\documentclass[journal,twoside,web]{ieeecolor}

\usepackage{jsen}
\usepackage{cite}
\usepackage{amsmath,amssymb,amsfonts}
\usepackage{algorithmic}
\usepackage{graphicx}
\usepackage{textcomp}
\usepackage{wrapfig}
\usepackage{url}
\usepackage{hyperref}
\usepackage{comment}
\usepackage{soul}
\usepackage{multirow}
\hypersetup{breaklinks=true}

\def\BibTeX{{\rm B\kern-.05em{\sc i\kern-.025em b}\kern-.08em
    T\kern-.1667em\lower.7ex\hbox{E}\kern-.125emX}}

\definecolor{abstractbg}{rgb}{0.89804,0.94510,0.83137}
\begin{document}

\title{A Miniaturized Wi-Fi HaLow-Based Wireless Imaging Transfer System for Capsule Endoscopy}

\author{Michele Bertozzi, Lorenzo Amelio, Sebastiano Strangio \IEEEmembership{Senior Member, IEEE}, Nikita J. Greenidge, Benjamin Calm\'e, Pietro Valdastri, \IEEEmembership{Fellow, IEEE}, Giuseppe Iannaccone \IEEEmembership{Fellow, IEEE}, \\ and Alessandro Catania \IEEEmembership{Senior Member, IEEE},
\thanks{This work has been submitted to the IEEE for
possible publication. Copyright may be transferred
without notice, after which this version may no
longer be accessible.}
\thanks{This work was supported by the EC Horizon 2020 Research and Innovation Programme under GA AUTOCAPSULE 952118 and partially supported by the Italian Ministry of University and Research (MUR) under the Forelab project of the ``Dipartimenti di Eccellenza'' programme.}
\thanks{Michele Bertozzi and Lorenzo Amelio are co-first authors.}
\thanks{Lorenzo Amelio, Sebastiano Strangio and Alessandro Catania are with the Department of Information Engineering, University of Pisa, Pisa, Italy (e-mail: l.amelio@studenti.unipi.it; sebastiano.strangio@unipi.it; alessandro.catania@unipi.it). Michele Bertozzi and Giuseppe Iannaccone are with the Department of Information Engineering, University of Pisa, Pisa, Italy, and with Clavis s.r.l. and Quantavis s.r.l., both located in Largo Padre Renzo Spadoni, Pisa, Italy (e-mail: michele.bertozzi@phd.unipi.it; giuseppe.iannaccone@unipi.it). Nikita J. Greenidge, Benjamin Calm\'e and Pietro Valdastri are with the STORM Lab, Institute of Autonomous Systems and Sensing (IRASS), School of Electronic and Electrical Engineering, University of Leeds, U.K. (e-mail: elnjg@leeds.ac.uk; b.p.calme@leeds.ac.uk; p.valdastri@leeds.ac.uk).}
}

\IEEEtitleabstractindextext{%
\fcolorbox{abstractbg}{abstractbg}{%
\begin{minipage}{\textwidth}%
\begin{wrapfigure}{r}{0.42\textwidth}
\vspace{-2mm}
\centering
\includegraphics[width=0.42\textwidth]{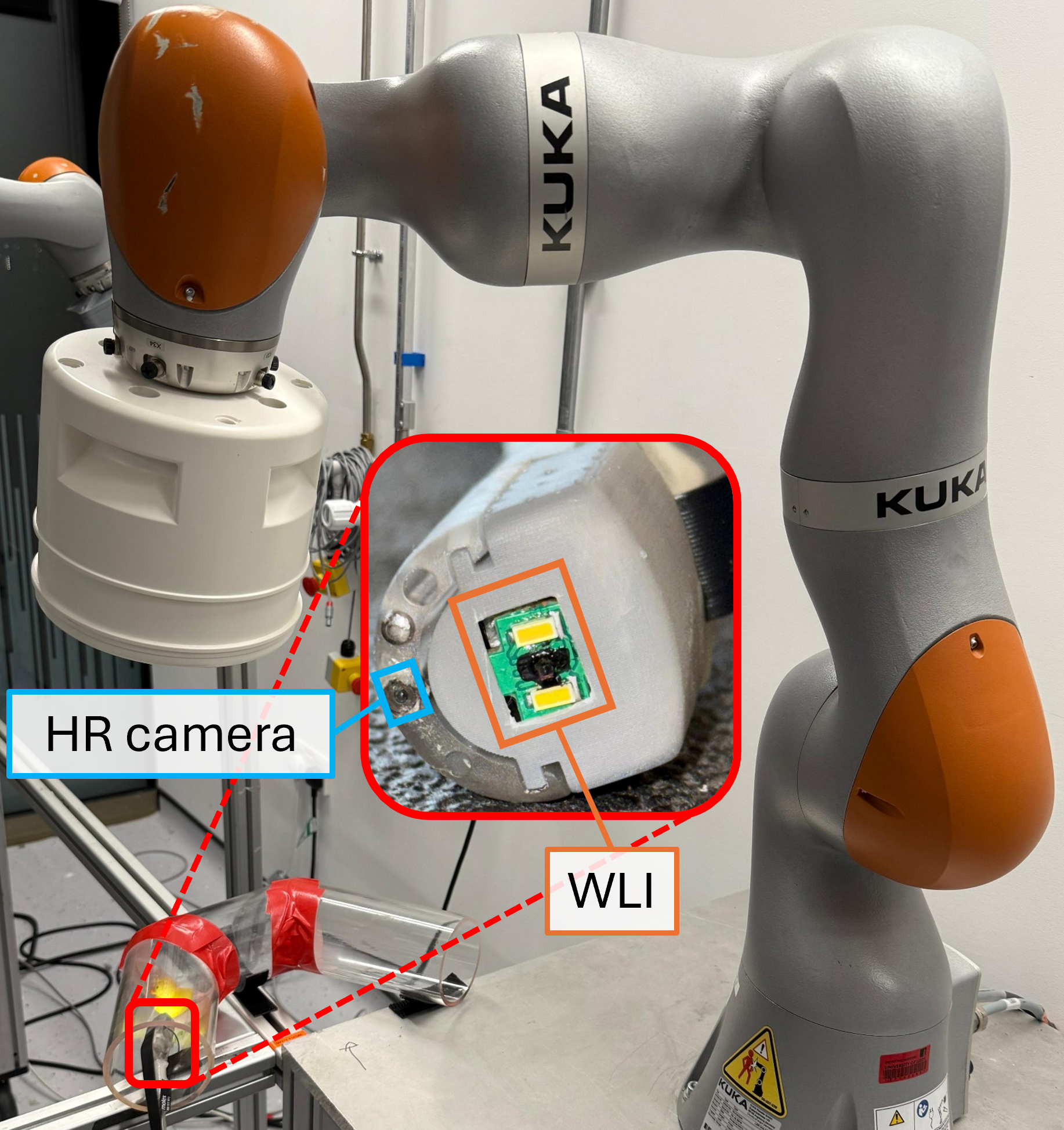}%
\vspace{-2mm}
\end{wrapfigure}

\begin{abstract}
Wireless capsule endoscopy is constrained by severe limitations in size, available power, and reliable data transmission through biological tissues. Although recent advances in miniature imaging sensors have improved image quality, the wireless transmission of color images at practical frame rates remains a challenging task under the tight capsule constraints.

We present the design and the experimental validation of a miniaturized Wireless Imaging Transfer (WIT) system for capsule endoscopy based on a sub-GHz IEEE 802.11ah (Wi-Fi HaLow) wireless link. In addition, a preliminary in-vivo validation of Wi-Fi HaLow communication is reported, showing that a stable through-body wireless link can be established across biological tissues in a clinically relevant porcine model. These in-vivo results provide direct experimental support for the adoption of sub-GHz Wi-Fi HaLow for capsule endoscopy applications.

Experimental results demonstrate stable wireless image transmission at a resolution of 320~$\times$~320 pixels, with measured end-to-end data rates above 12 Mbps and a transmission energy efficiency of 20.7~nJ/bit. We also report a detailed power consumption analysis, with a breakdown of the contributions from the imaging sensor, illumination, digital processing, and RF transmission. When all active subsystem contributions are considered, including the camera, digital processing, RF transmission, and illumination at 60\% LED brightness, the total power consumption is approximately 477~mW, corresponding to a complete-chain energy consumption of approximately 39.6~nJ/bit. Overall, the measured performance confirms that sub-GHz Wi-Fi HaLow provides a promising trade-off among throughput, robustness, and energy efficiency for wireless imaging in capsule endoscopy.

%Overall, the reported results indicate that Wi-Fi HaLow is a viable solution for through-body data communication and that the proposed WIT architecture represents a scalable platform for real-time color imaging in future autonomous capsule endoscopy systems.
\end{abstract}

\begin{IEEEkeywords}
sub-GHz communication,
Wi-Fi HaLow (IEEE 802.11ah),
low-power embedded systems,
through-body communication,
wireless capsule endoscopy,
wireless data transmission
\end{IEEEkeywords}

\end{minipage}%
}%
}%
\maketitle

%========================
% I. INTRODUCTION
%========================
\section{Introduction}
\label{sec:introduction}

\IEEEPARstart{I}{ntra-body}, implantable, and ingestible medical devices are progressively redefining the paradigm of in-body diagnostics by enabling continuous or on-demand monitoring of physiological parameters without the need for tethered instrumentation. Among these devices, wireless capsule endoscopy represents one of the most established and clinically relevant examples, providing minimally invasive access to the gastrointestinal (GI) tract \cite{Iddan2000, Goenka2014}.
By integrating imaging, sensing, and data transmission within a swallowable platform, capsule systems have extended diagnostic reach beyond conventional flexible endoscopy.

%%%%%%%%%%%%%%%%%%%%%%%%%%%%%%%%%%%%%%%%%%%%%%%%%%%%%%%%%%%%%%%%%%%%%%%%%%%%%%%%
% TABLE — WIRELESS PROTOCOL COMPARISON
%%%%%%%%%%%%%%%%%%%%%%%%%%%%%%%%%%%%%%%%%%%%%%%%%%%%%%%%%%%%%%%%%%%%%%%%%%%%%%%%
\begin{table*}[!t]
\centering
\caption{Comparison of wireless protocols.}
\label{tab:protocol_comparison}
\renewcommand{\arraystretch}{1.3}
\begin{tabular}{lcccc}
\hline
\textbf{Protocol} &
\textbf{Frequency} &
\textbf{Bandwidth} &
\textbf{Max. Data Rate} &
\textbf{Range} \\
\hline
\hline
HBC \cite{IEEE802156}         & 16-27MHz          & 5.25 MHz & 1.32 Mbps & Limited \\
Wi-Fi HaLow \cite{WiFiAlliance2021HaLow}& \multirow{4}{*}{sub-GHz} & 1-16 MHz & 86.7 Mbps \cite{WiFiAlliance2021HaLow}& $>$1 km \\
Wi-SUN \cite{WiFiAlliance2021HaLow}     &                         & 100-1200 kHz & 800 kbps & $<$1 km \\
ZigBee  \cite{WiFiAlliance2021HaLow}    &                         & $\sim$600 kHz & 40 kbps & $<$100 m \\
LoRaWAN \cite{WiFiAlliance2021HaLow}    &                         & 125-500 kHz & 27 kbps & $<$10 km \\
BLE   \cite{BLE}        & 2.4 GHz & 2 MHz & 2 Mbps & $<$100 m \\
Wi-Fi 6/6E \cite{IEEE80211ax2021} & 2.4/5/6GHz                   & 20-160 MHz & $>$600 Mbps & $<$100 m \\
UWB \cite{Tsang2005}         & 3.1-10.6 GHz           & $\sim$500 MHz & $>$100 Mbps & $<$10 m \\
\hline
\end{tabular}
\end{table*}
\begin{comment}
\begin{table*}[!t]
\centering
\renewcommand{\arraystretch}{1.1}
\caption{Energy efficiency comparison of wireless links used for high-data-rate biomedical or IoT applications. Energy-per-bit values are either reported in the literature or derived from the reported power consumption and data rate.}
\label{tab:energy_bit_comparison}
\begin{tabular}{lccccc}
\hline
\textbf{Reference / Device} & \textbf{Protocol} & \textbf{Data Rate} & \textbf{Power} & \textbf{Energy/bit} & \textbf{Type} \\
\hline
This work & Wi-Fi HaLow & $\sim$12 Mbps & $\sim$250 mW & 20.7 nJ/bit & Commercial module \\

\cite{Gao2011} & UWB & 10 Mbps & 3.5 mW (TX) & 0.35 nJ/bit & Custom ASIC \\

\cite{Park2018} & Human Body Comm. & 6 Mbps & 3.7 mW (TX) & 0.62 nJ/bit & Custom ASIC \\

nRF52840 \cite{nRF52840Datasheet} & BLE 5 & 2 Mbps & $\sim$19 mW (TX) & $\sim$9.6 nJ/bit & Commercial SoC \\

ESP32-C3 \cite{ESP32C3Datasheet} & Wi-Fi 4 & 54 Mbps & $\sim$0.94 W (TX) & $\sim$17 nJ/bit & Commercial SoC \\

\hline
\end{tabular}
\end{table*}
\end{comment}

Wireless capsule endoscopy (WCE) is a noninvasive clinical procedure that provides an extensive evaluation of the digestive system through a multi-sensor device.
The capsule-scale device may embed different subsystems such as White-Light Imaging (WLI), physiological sensors (pH, temperature, pressure) \cite{SmartPill}
and magnetic localization units, all competing for the same energy and power budget \cite{Ciuti2011}, limited by a small battery or, more recently, by a wireless power transfer (WPT)
system, which is often envisioned as a complementary technology to enable continuous operation \cite{Catania2025}.

While low-latency streaming is not strictly required, adequate temporal and spatial resolution are essential to maximize the diagnostic yield.
Higher frame rates increase the number of spatial samples acquired, thereby reducing the probability of missed lesions, particularly in fast
peristaltic phases. Providing high image quality is particularly energy-demanding, resulting in a trade-off between image fidelity and total examination duration.
Therefore, an energy-efficient wireless link becomes a key enabler for effective image transmission without compromising the limited energy budget of the capsule.

Commercial endoscopic capsules typically rely on proprietary low-data-rate links combined with on-board compression to reduce bandwidth requirements \cite{Iddan2000}. 
While this approach enables battery-powered operation, it increases architectural complexity and may introduce compression artifacts that degrade image quality \cite{Alam2017}. 
The limited transparency of proprietary physical layers further complicates reproducibility and comparative research.

In addition, the communication channel must propagate through highly lossy and dispersive layers of biological tissue (e.g., skin, adipose tissue, muscle, and intestinal walls),
whose frequency-dependent dielectric permittivity and conductivity introduce a fundamental trade-off among carrier frequency, antenna size, and tissue attenuation \cite{Cao2024}. The choice of the carrier frequency is crucial for optimizing the performance of the capsule in terms of data rate, energy efficiency, and link robustness, and therefore is a system-level design parameter tightly coupled with the overall energy budget \cite{Yuce2012}.

Widely adopted standard wireless communication protocols are summarized in Table~\ref{tab:protocol_comparison}. 
Bluetooth Low Energy (BLE) at 2.4 GHz offers lower power consumption than Wi-Fi and broad ecosystem support, but is limited to sub-Mbps data rates and suffers
from high attenuation in tissue: while BLE supports physical-layer data rates up to 2 Mbps, its effective application-layer throughput can be substantially lower under in-body operating conditions. Zhou et al. \cite{Zhou2026} demonstrated, in water, a maximum throughput below 1 Mbps using BLE and around 8 Mbps using Wi-Fi at 2.4 GHz and 5 GHz. The calculated best case (zero distance between transmitter and receiver) energy per bit for BLE is around 22 nJ/bit, while for Wi-Fi it is around 31.25 nJ/bit. Moreover, a major advantage of high-frequency communication lies in the
small form factor of the antennas, which can be easily integrated into capsules.

In contrast, lower frequency carriers, adopted in the Human-Body Communication (HBC) defined standard IEEE 802.15.6, have the opposite characteristics.
Their improved propagation through tissues is hindered by the reliance on larger electronics,
which reduces the available space inside the capsule. Moreover, it supports data rates up to approximately 1.3 Mbps which are suitable for low-data-rate body-area communications, but insufficient for the uncompressed imaging stream targeted here\cite{Cao2024}.

Ultra-wideband (UWB) solutions can support large bandwidths (minimum of 500 MHz) in the range of 3.1 to 10.6 GHz but require complex RF front-ends and may experience stronger absorption at higher frequency in conductive media \cite{Kshetrimayum2009}. Despite the resulting lower link budget, the high achievable throughput has made UWB a popular choice
for wireless imaging, and several groups developed antennas for operation at these high frequencies \cite{BenWang2024}. 
A notable design trade-off concerns the adoption of sub-GHz carrier frequencies,
commonly used in long-range IoT systems such as LoRaWAN, NB-IoT and ZigBee.
These technologies typically operate with channel bandwidths of a few hundred kHz (or below), achieving high link budgets at the expense of spectral
efficiency. While suitable for low-data-rate applications, they are not ideal for high-resolution imaging,
which requires multi-megabit-per-second links to sustain practical frame rates without excessive compression.

In this context, IEEE 802.11ah (Wi-Fi HaLow) has emerged as a sub-GHz
alternative combining extended range with multi-megabit physical-layer
capability \cite{Adame2014}. The standard supports channel bandwidths from
1 to 16 MHz and data rates ranging from approximately 150 kbps up to
86.7 Mbps in single-stream configurations. By exploiting OFDM modulation
with adaptive MCS selection, Wi-Fi HaLow offers a favorable trade-off
between penetration depth and spectral efficiency compared to
2.4 GHz technologies, while maintaining throughput compatible with
wireless imaging applications. In practical use, data rates on the order of tens of Mbps can be achieved with a 4 MHz bandwidth and a 64-QAM modulation scheme \cite{WiFiAlliance2021HaLow}. Despite this potential, no prior work has experimentally validated a Wi-Fi HaLow link through living biological tissues or under in-vivo conditions. IEEE 802.11ah has previously been investigated for healthcare monitoring \cite{Wikanta2020, Taha2024}; however, the available studies are primarily based on analytical path-loss models. More generally, experimental demonstrations of fully miniaturized, capsule-compatible WIT systems validated in tissue-equivalent
environments remain limited, and quantitative subsystem-level power analyses are rarely reported in the literature \cite{Chen2009}.

This work leverages the advantages of IEEE 802.11ah to design and experimentally validate a miniaturized WIT system. The proposed architecture integrates a compact color imaging front-end based on a NanEyeC camera module (ams OSRAM \cite{amsOSRAM_NanEyeC}), a low-power STM32U575 microcontroller (STMicroelectronics \cite{STMicroelectronics_STM32U575OI}) for data buffering and handling, and an SX-SDMAH Wi-Fi HaLow transceiver module (Silex Technology \cite{Silex_SX_SDMAH}) for wireless transmission. The complete imaging chain is implemented on capsule-compatible hardware and validated in a colon phantom using the oloid-based magnetic actuation platform described in \cite{Greenidge2025}. The platform employs an external permanent magnet mounted on a KUKA LBR Med R820 robotic manipulator to control the position and orientation of the magnetically coupled capsule.

This work provides three main contributions: (i) a preliminary in-vivo validation of Wi-Fi HaLow communication through biological tissues, demonstrating the feasibility of stable through-body data transmission; (ii) an experimentally validated end-to-end miniaturized WIT chain sustaining data rates above 12 Mbps; and (iii) the benchmarking of the complete imaging chain implemented on capsule-compatible hardware in laboratory conditions, under manipulation with the oloid-based magnetic actuation platform, against a wired Olympus PCF-160AL high-resolution reference colonovideoscope. In addition, a detailed current-consumption analysis is reported for both the wireless transmission subsystem and the imaging chain, enabling explicit estimation of the corresponding power contributions.

%===========================================================
% II. ELECTROMAGNETIC CONSTRAINTS AND LINK DESIGN RATIONALE
%===========================================================
\section{Electromagnetic Constraints and Link Design Rationale}
\label{sec:em_rationale}
The selection of a communication protocol operating around 900 MHz is primarily driven by a trade-off between achievable throughput and propagation loss.
Lower-frequency carriers are typically associated with narrower allocated bandwidths and consequently reduced throughput,
whereas higher-frequency bands offer wider channels at the cost of increased propagation losses and hence, a reduced link budget.

Secondly, in-body wireless links are fundamentally constrained by the dispersive
electromagnetic properties of biological tissues.
In the frequency range of interest for capsule communication,
GI tissues can be modelled as lossy dielectrics characterized by
frequency-dependent relative permittivity $\varepsilon_r(f)$ and
conductivity $\sigma(f)$.
These parameters directly determine attenuation, achievable link margin,
and the transmit-power levels required to maintain reliable communication.

To quantify the impact of carrier frequency on propagation,
we evaluate the penetration depth $\delta$, a measure of how deeply an electromagnetic radiation can penetrate into a material.
It is defined as the inverse of the attenuation constant $\alpha$ of a plane wave propagating in a lossy medium:
\begin{equation}
\delta = \frac{1}{\alpha}.
\label{eq:delta_def}
\end{equation}
For a non-magnetic medium ($\mu \approx \mu_0$),
the attenuation constant can be expressed as:
\begin{equation}
\alpha =
2\pi f
\sqrt{\frac{\mu_0 \varepsilon_0 \varepsilon_r}{2}}
\sqrt{
\left[
\sqrt{1 +
\left(
\frac{\sigma}{2 \pi f \varepsilon_0 \varepsilon_r}
\right)^2}
- 1
\right]
},
\label{eq:alpha_full}
\end{equation}
where 
%$\omega = 2\pi f$,
$\varepsilon_0$ is the vacuum permittivity,
and $\sigma$ is the electrical conductivity.
Frequency-dependent dielectric parameters were taken from tabulated dispersive datasets
available in literature and in public databases (e.g., \cite{Gabriel1996, ITIS_database}).

\begin{figure}[!t]
    \centering
    \includegraphics[width=\columnwidth]{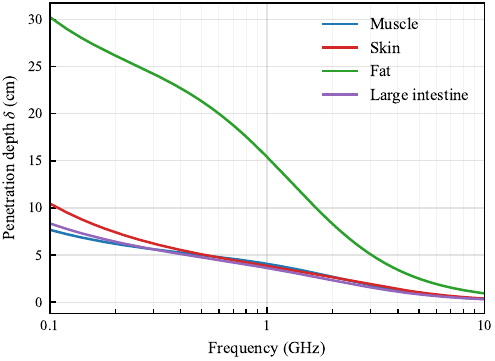}
    \caption{Electromagnetic penetration depth $\delta$
    in representative biological tissues over the 0.1-10 GHz range,
    computed from frequency-dependent dielectric properties.}
    \label{fig:penetration_depth100M_10GHz}
\end{figure}

Fig.~\ref{fig:penetration_depth100M_10GHz} shows that penetration depth
decreases monotonically with frequency.
This trend is driven by conductive losses, captured by the term
$\sigma/(f \varepsilon_r)$ in \eqref{eq:alpha_full}.
Highly hydrated tissues (e.g., muscle and large intestine)
exhibit stronger attenuation due to higher conductivity,
whereas low-conductivity tissues (e.g. fat) yield larger penetration depth.

In the sub-GHz region, $\delta$ remains on the order of several centimetres
for most tissues, whereas it rapidly decreases in the 2--3 GHz range.
Therefore, carrier frequencies around 0.9 GHz offer a favorable trade-off:
attenuation is significantly lower than at 2.4 GHz,
while data rate is not limited by narrow bandwidths. In addition, antenna size remains compatible with capsule-scale integration.
For practical link-budget reasoning, the attenuation constant can be
approximated in conductive media as:
\begin{equation}
\alpha \approx \sqrt{\pi f \mu \sigma},
\label{eq:alpha_approx}
\end{equation}
which yields attenuation values on the order of a few dB/cm around 900 MHz,
depending on tissue composition and hydration level
\cite{Gabriel1996, ITIS_database}.

Within this framework, adopting IEEE 802.11ah at sub-GHz frequencies is
motivated by two coupled factors:
\begin{itemize}
    \item Reduced tissue attenuation increases link budget, enabling more reliable communications than higher-frequency alternatives.
    \item Higher throughput is achievable compared to other protocols using the same or lower frequencies.
\end{itemize}
The experimental results reported in Section~\ref{sec:results} confirm that
the communication exhibited negligible packet loss and a throughput of 12--13 Mbps, compatible with the application,
thus supporting the selection of IEEE 802.11ah for capsule imaging.

%========================
% III. SYSTEM ARCHITECTURE OF THE WIT MODULE
%========================
\section{In-Vivo Validation of Wi-Fi HaLow}
\label{sec:in-vivo}

\begin{figure*}[!t]
    \centering
    \includegraphics[width=\textwidth]{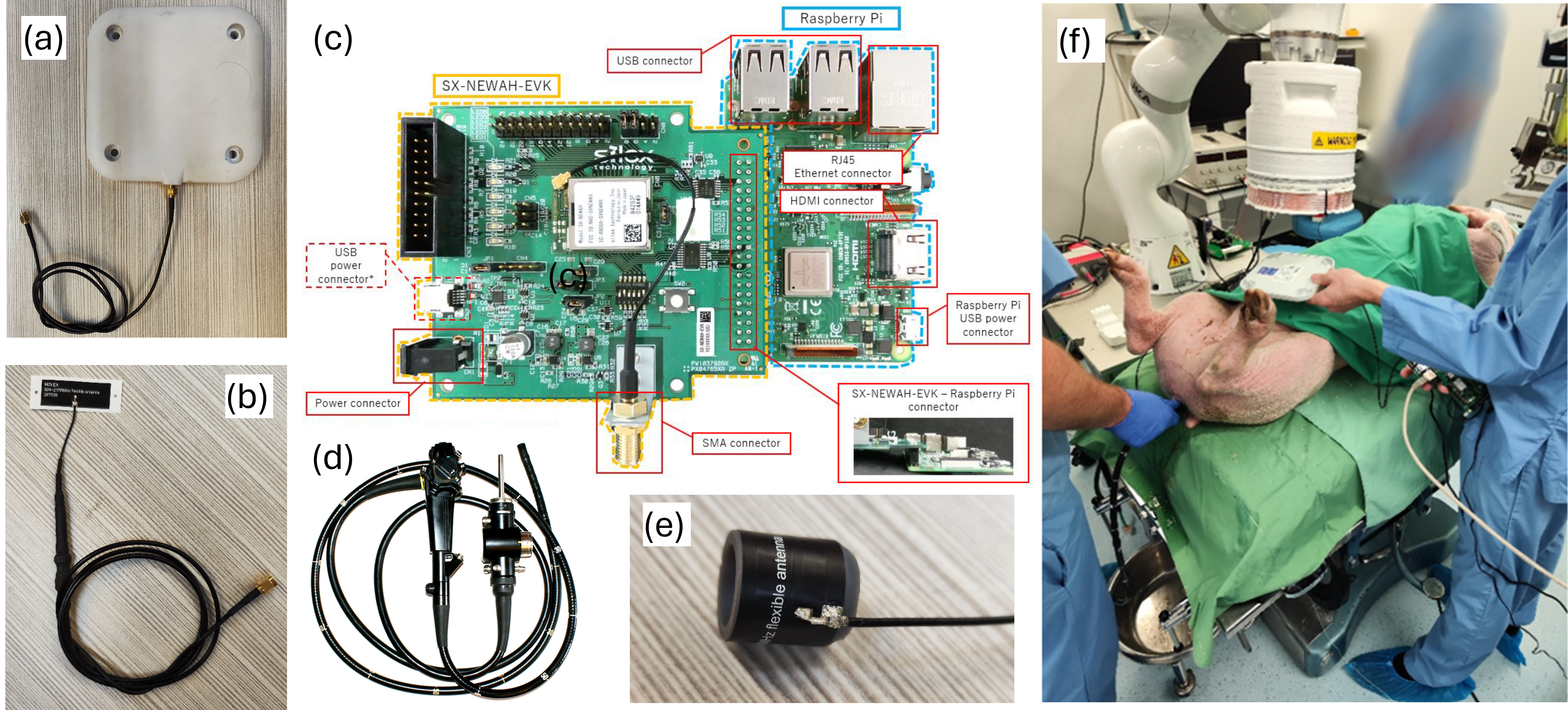}
    \caption{Experimental setup for the in-vivo validation of the Wi-Fi HaLow data transfer system. (a) Circularly polarized antenna connected to the external AP. (b) Flexible antenna connected to the STA and used for internal placement. (c) SX-NEWAH-EVK evaluation board interfaced with a Raspberry Pi for system control and data handling, used for both AP and STA. (d) Standard colonoscope (Olympus 100 series) employed during the procedure. (e) Custom 3D-printed add-on mounted at the endoscope tip to host the flexible antenna. (f) In-vivo experimental setup during the validation campaign.}
    \label{fig:in-vivo_setup}
\end{figure*}

\normalcolor

Before the development of the fully miniaturized WIT prototype presented in this work, an in-vivo validation campaign was carried out to assess the feasibility of IEEE 802.11ah communication through biological tissues under realistic operating conditions. All animal procedures were performed in compliance with the Animals (Scientific Procedures) Act 1986 and the relevant NC3Rs and ARRIVE guidelines. To provide a clinically meaningful validation scenario, a porcine model was selected because of its close anatomical similarity to the human gastrointestinal tract. These experiments were conducted at an earlier stage of the project, before the design of the WIT system described in this paper, with the specific aim of verifying the robustness of the Wi-Fi HaLow protocol for through-body communication under realistic in-vivo conditions.

At that stage, a capsule-integrable implementation of the wireless subsystem was not yet available, and the objective was therefore to validate the communication channel independently of the final hardware integration. For this reason, the in-vivo campaign was intentionally designed to test the wireless link and antenna placement in the animal while keeping the Wi-Fi HaLow transceiver electronics outside the body. Consistently with this preliminary validation purpose, the experiments were carried out using a different and earlier Wi-Fi HaLow transceiver platform, namely the SX-NEWAH developed by Silex Technology \cite{Silex_SX_NEWAH} and based on the Newracom NRC7292 system-on-chip \cite{Newracom_NRC7292}. In contrast, the WIT platform presented in this article adopts the more recent SX-SDMAH module from the same manufacturer, which represents an evolution of the previous solution and is based on the more advanced Morse Micro MM6108 system-on-chip. Therefore, the in-vivo tests reported in this section should be regarded as an initial validation of the communication principle and of the through-body channel, whereas the WIT architecture discussed in the remainder of the paper represents the subsequent hardware evolution toward a more integrated and higher-performance implementation.

The experimental setup is shown in Fig.~\ref{fig:in-vivo_setup}. The external side of the link was implemented through an SX-NEWAH-EVK platform interfaced with a Raspberry Pi, shown in Fig.~\ref{fig:in-vivo_setup}(c), configured as Access Point (AP) and connected to a Times-7 A5020 circularly polarized antenna, reported in Fig.~\ref{fig:in-vivo_setup}(a), positioned over the animal abdomen during the experiment. On the internal side, a second SX-NEWAH-EVK board, again of the type shown in Fig.~\ref{fig:in-vivo_setup}(c), was configured as Station (STA). Its RF port was connected through a 914 mm RG-174 coaxial cable with SMA connector to a flexible Molex 207235 antenna, shown in Fig.~\ref{fig:in-vivo_setup}(b), which was mounted at the tip of a standard Olympus 100 series colonoscope [Fig.~\ref{fig:in-vivo_setup}(d)] by means of a custom 3D-printed add-on specifically designed to host the antenna at the endoscope tip [Fig.~\ref{fig:in-vivo_setup}(e)]. Although this tethered configuration is more penalizing than the final use case, since cable losses and mismatch can degrade the effective link performance, it enabled an early and controlled validation of through-body wireless communication before the availability of a fully integrated capsule-scale prototype. The overall in-vivo arrangement during the validation campaign is shown in Fig.~\ref{fig:in-vivo_setup}(f).

During the in-vivo validation, the wireless link was assessed in terms of application-layer throughput and RF front-end power consumption. Measurements were performed using iPerf in UDP mode to evaluate a Wi-Fi HaLow connection at 915 MHz using a 4 MHz channel bandwidth, while the current absorbed by the SX-NEWAH transceiver platform was recorded simultaneously. The capsule antenna was placed at a nominal distance of approximately 15 cm from the external receiver. Over the relatively small distance variations occurring during capsule manipulation, no appreciable change in link performance was observed, confirming the robustness of the adopted Wi-Fi HaLow configuration under the investigated conditions. Depending on the selected transmit-power setting by means of the programmable parameter \textit{txp}, the measured throughput ranged from 1.8 Mbps to 5.3 Mbps, with an average current consumption between 54 mA and 126 mA from a 3.3 V supply, corresponding to 178–415 mW. Although the absolute values depend on the specific configuration of the adopted commercial evaluation kit, the results clearly demonstrate that a Wi-Fi HaLow link can be reliably established through biological tissues in a realistic in-vivo scenario. 

\begin{figure}[!t]
    \centering
    \includegraphics[width=0.9\columnwidth]{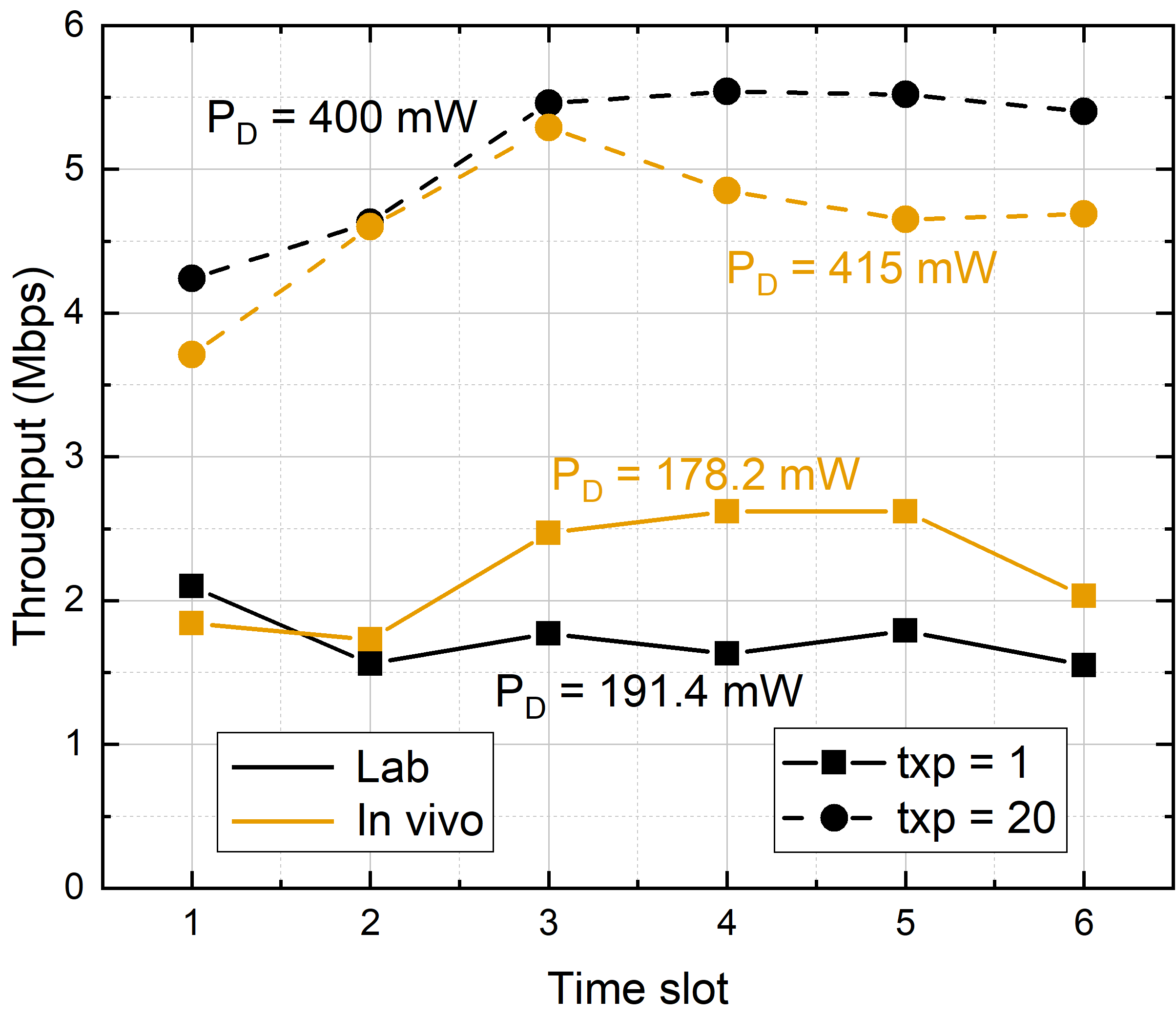}
    \caption{Comparison of the application-layer throughput measured with the SX-NEWAH Wi-Fi HaLow platform under laboratory and in-vivo conditions, for two transmit-power settings selected through the programmable parameter txp. The average dissipated power associated with each operating condition is also indicated.}
    \label{fig:comp_in-vivo_in-lab}
\end{figure}

Most importantly, Fig.~\ref{fig:comp_in-vivo_in-lab} shows that the in-vivo results are overall comparable with the corresponding laboratory measurements obtained with the same SX-NEWAH platform under similar operating conditions. Although slight differences can be observed in both throughput and dissipated power, these variations can also be ascribed to small differences in link distance and to the adaptive selection of transmission parameters, such as modulation and coding scheme and channel mapping, by the commercial transceiver platform. Nevertheless, the measured throughput remains consistently in the multi-Mbps range in both cases, confirming that propagation through biological tissues does not prevent stable Wi-Fi HaLow communication and providing a direct experimental validation of through-body data transmission.

At the same time, the relatively high power consumption observed in these tests is mainly related to the specific transceiver generation adopted in this preliminary setup. For this reason, the miniaturized WIT prototype presented in this work employs the more recent SX-SDMAH module based on the Morse Micro MM6108, which enables a more efficient implementation and higher achievable data rates. In this sense, the in-vivo campaign should be regarded as a preliminary proof of communication feasibility, whereas the WIT architecture described in the following sections represents its more advanced technological evolution.

%========================
% IV. SYSTEM ARCHITECTURE OF THE WIT MODULE
%========================
\section{System Architecture of the WIT Module}
\label{sec:architecture}

The proposed WIT module implements a complete capsule-side acquisition and wireless streaming chain.
The architecture is designed to satisfy three simultaneous constraints:
compact form factor, reduced average power consumption, and sustained multi-megabit throughput.

\begin{figure}[!t]
    \centering
    \includegraphics[width=\columnwidth]{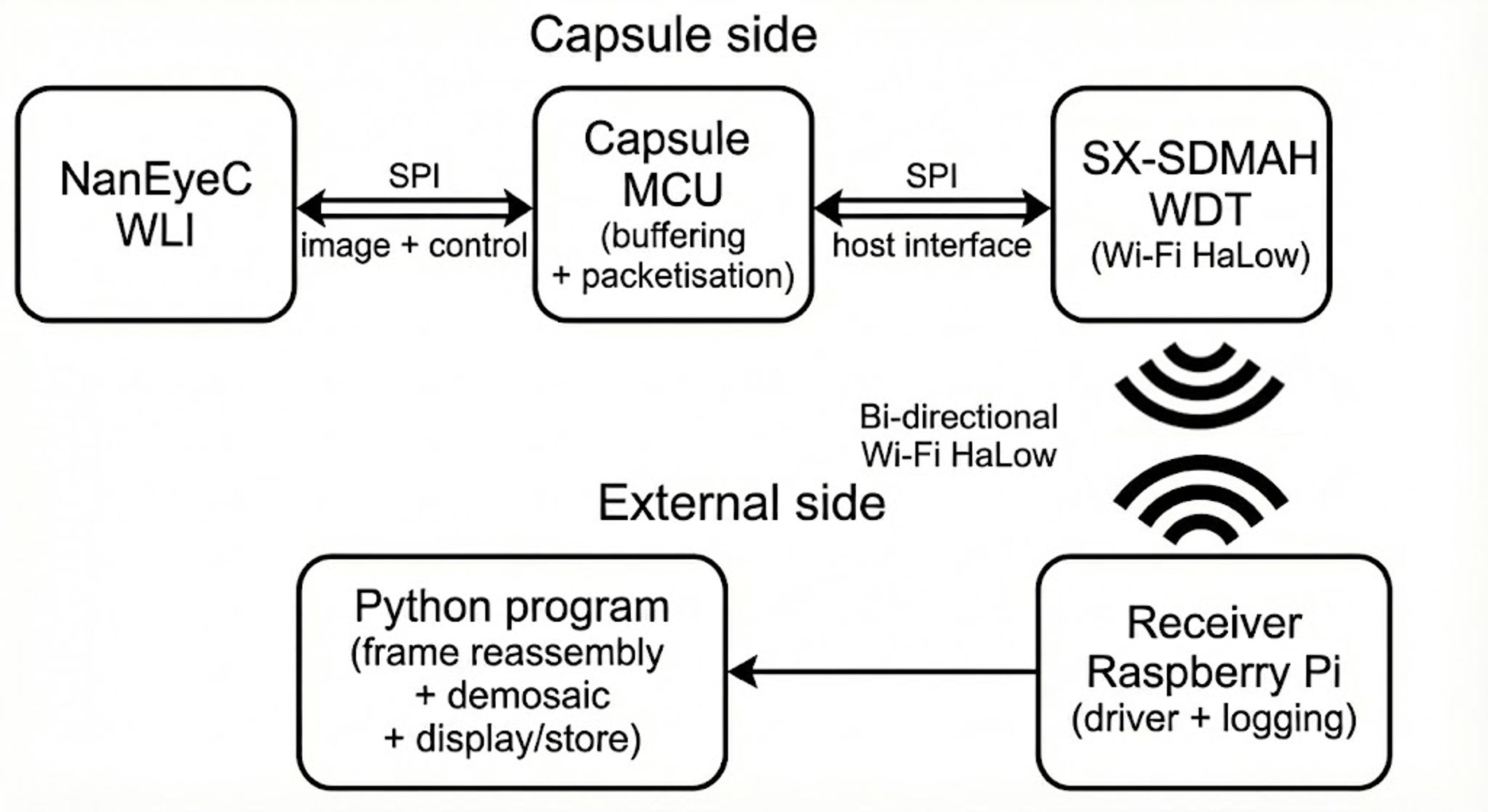}
    \caption{Block diagram of the proposed WIT module, illustrating the main subsystems and their interconnections.}
    \label{fig:BD}
\end{figure}

At the system level (Fig.~\ref{fig:BD}), the WIT chain comprises: (i) a miniaturized WLI front-end, based on the NanEyeC camera, for image acquisition at the capsule tip; (ii) a low-power microcontroller for data acquisition, buffering, packetization, and local control; and (iii) a sub-GHz IEEE~802.11ah transceiver for wireless transmission through the Wi-Fi HaLow link. On the external side, an access point interfaced with a Raspberry Pi is used to receive the transmitted packets, reconstruct the image frames, and handle their visualization and storage. In addition to the forward image-transmission path, the system also supports a reverse control path, enabling the external unit to send commands back to the capsule-side electronics for the adjustment of operating parameters such as exposure time and LED intensity.

\begin{figure}[!b]
    \centering
    \includegraphics[width=\columnwidth]{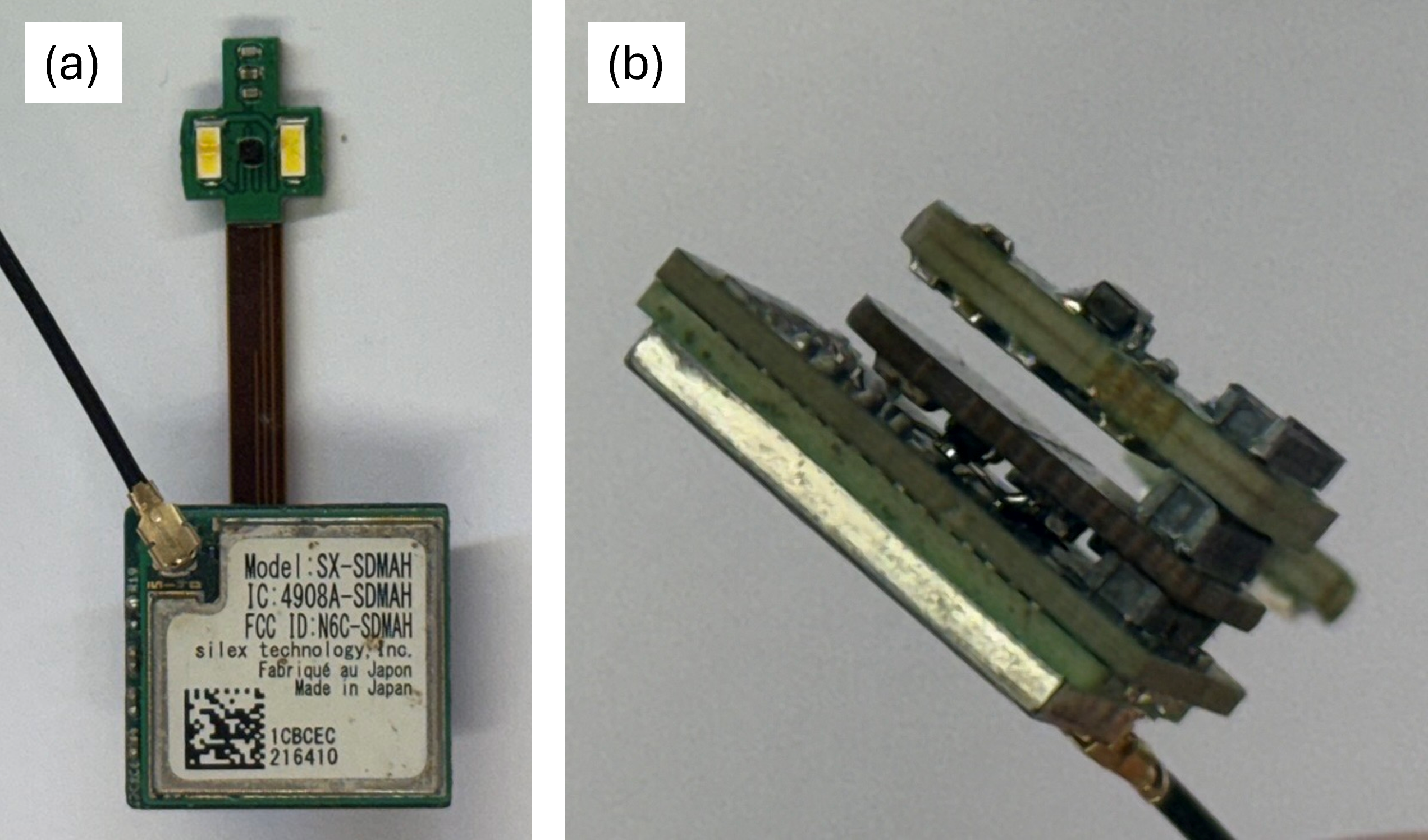}
    \caption{System-level architecture of the proposed WIT module. The capsule-side electronics integrate
    the imaging front-end, MCU-based data handling, and
    IEEE 802.11ah sub-GHz transceiver.}
    \label{fig:WIT_System}
\end{figure}

Fig.~\ref{fig:WIT_System} illustrates the system-level architecture, composed of a total of three PCBs, used to assemble the subsystems listed above.

%------------------------
\subsection{Imaging Front-End}
\label{subsec:arch_imaging}

The imaging subsystem is based on the low-power NanEyeC miniature CMOS color camera module, featuring an ultra-compact form factor
of $1\,\mathrm{mm} \times 1\,\mathrm{mm} \times 2\,\mathrm{mm}$, which places it among the smallest commercially available fully integrated color camera modules.
The sensor provides a native resolution of $320 \times 320$ pixels and supports programmable exposure times, allowing operation under low-light conditions typical of in-body scenarios.

A key architectural advantage of the NanEyeC lies in its single-ended digital serial output interface, which enables direct connection to a
low-power microcontroller without intermediate deserialization or signal-conditioning stages. This removes the need for additional interface electronics,
thereby reducing PCB area, routing complexity, and static power consumption.

\begin{figure}[!t]
    \centering
    \includegraphics[width=\columnwidth]{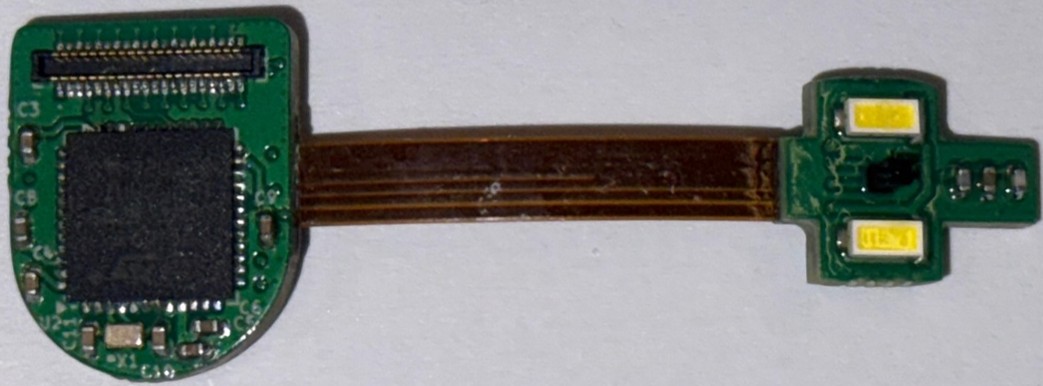}
    \caption{Miniaturized WLI front-end based on
    the NanEyeC camera and integrated illumination LEDs
    implemented on a rigid-flex PCB.}
    \label{fig:MiniaturizedWLI}
\end{figure}

The camera and the illumination LEDs are integrated on a custom rigid-flex PCB, as shown in Fig.~\ref{fig:MiniaturizedWLI}.
The flexible section routes signals from the distal tip, where the optics and LEDs are placed, to the rigid section hosting the microcontroller.
Pulse-width modulation (PWM) control is implemented to regulate LED brightness, enabling dynamic trade-offs between illumination quality
and average power consumption.

The power supply of the NanEyeC imaging module is managed through a dedicated NanEyeC Enable Board. The board integrates a low-dropout linear regulator (LDO)
with enable, allowing the control of the camera power state by firmware and therefore preventing desynchronization issues observed in earlier prototypes.

%------------------------
\subsection{Microcontroller and Data Handling}
\label{subsec:arch_mcu}

Image acquisition and data management are performed by an STM32U575 microcontroller.
The device was selected for its combination of ultra-low-power operating modes, sufficient processing capability, and integrated high-speed serial peripherals.
The STM32U575CIU6Q variant, packaged in a 48-pin UFQFPN (7\,mm $\times$ 7\,mm), was selected to reduce routing complexity and occupied PCB area within the constrained capsule form factor.

The firmware implements a lightweight acquisition pipeline: raw data are received via DMA, stripped of non-informative protocol bits (e.g. Start/Stop bits, End of Frame/Row bytes), and then sent to the transceiver through a full-duplex SPI interface.
In addition, the same microcontroller manages bidirectional control, allowing remote adjustment of exposure time and LED intensity through the same wireless link.
The MCU is mounted on the same rigid-flex PCB of the camera module, specifically designed for capsule-scale integration.

%------------------------
\subsection{Sub-GHz IEEE 802.11ah Transceiver}
\label{subsec:arch_wireless}
\normalcolor

For the proposed WIT prototype, we adopted the latest available SX-SDMAH IEEE 802.11ah (Wi-Fi HaLow) module, whose MM6108-based architecture provides higher throughput and improved power efficiency compared with the earlier SX-NEWAH platform used for the in-vivo feasibility tests, which had already demonstrated reliable Wi-Fi HaLow communication through biological tissues. 
The SX-SDMAH module is based on the Morse Micro MM6108 system-on-chip \cite{MorseMicro_MM6108}, which integrates the RF front-end (PA, LNA and matching network), baseband processor, and full MAC-layer functionality compliant with IEEE 802.11ah. It supports channel bandwidths from 1 to 16 MHz and multiple MCS levels, enabling adaptive data-rate operation according to link conditions. The device is interfaced to the STM32U575 via a high-speed SPI connection, allowing the MCU to handle frame buffering while the transceiver autonomously manages modulation, channel access, framing, and retransmissions.
A flexible sub-GHz antenna (MOLEX 211140-0100 \cite{Molex}) is employed to accommodate the constrained capsule geometry while maintaining acceptable radiation efficiency.

%\begin{figure}[!t]
%    \centering
%    \includegraphics[]{figures/MiniaturizedWDT.jpg}
%    \caption{Miniaturized IEEE 802.11ah transceiver module
%    mounted on a capsule-compatible PCB.}
%    \label{fig:MiniaturizedWDT}
%\end{figure}

Although the radio module represents the largest single component in the current prototype, its architecture allows further integration at SoC level,
potentially reducing footprint and parasitic losses. The adoption of the commercial module was primarily motivated by the need to experimentally
demonstrate the feasibility of the selected wireless protocol as a proof of concept prior to further miniaturization efforts.
%------------------------
\begin{table}[!h]
\centering
\caption{Physical Dimensions of the Miniaturized WIT Subsystems}
\label{tab:dimensions}
\begin{tabular}{lccc}
\hline
Board & Length [cm] & Width [cm] & Height [cm] \\
\hline
WLI Board & 1.25 & 1.5 & 0.20 \\
NanEyeC Enable Board & 1.25 & 1.5 & 0.25 \\
WDT Board & 1.8 & 1.9 & 0.40 \\
\hline
\end{tabular}
\end{table}

\subsection{Miniaturized Stack Implementation}
\label{subsec:arch_stack}

\begin{figure}[!t]
    \centering
    \includegraphics[width=\columnwidth]{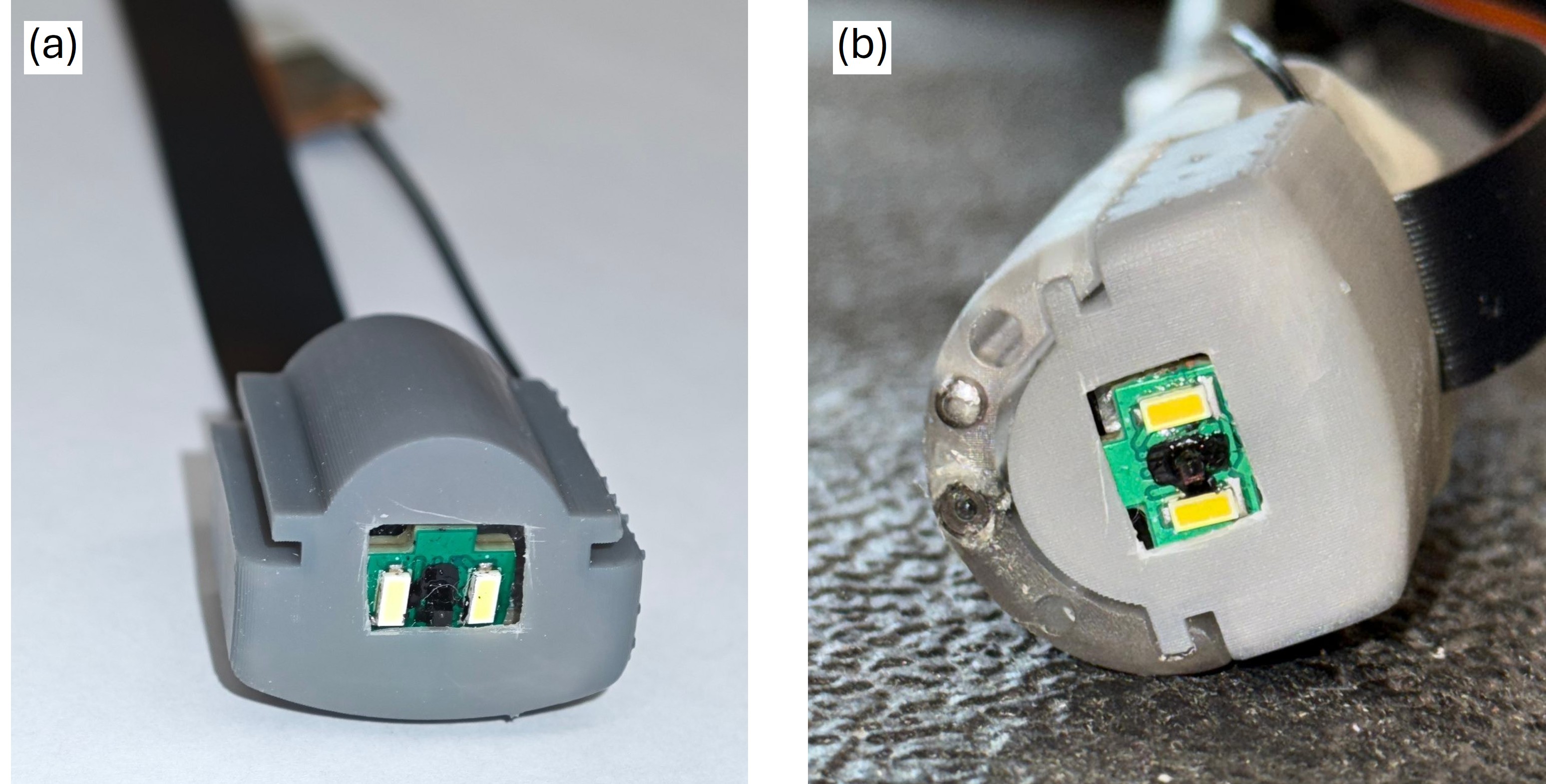}
    \caption{Prototype integration. (a) Dedicated compartment housing the fully miniaturized WIT stack. (b) Complete oloid-shaped capsule used in the experimental tests, integrating the WIT compartment, an internal permanent magnet for magnetic actuation, and the wired reference camera used for benchmarking.
    }
    \label{fig:compartment}
\end{figure}

Table~\ref{tab:dimensions} summarizes the physical dimensions of the miniaturized custom PCBs composing the WIT stack.
The imaging front-end (WLI board) occupies the smallest footprint and is positioned at the distal tip of the capsule,
while the WDT board hosting the IEEE 802.11ah module represents the most volumetrically demanding subsystem.
The total stacked height remains below 1 cm, which is compatible with capsule-scale integration
when combined with mechanical housing and additional subsystems.

The complete WIT module is realized as a stacked assembly of miniaturized PCBs.
The rigid-flex imaging board is positioned at the distal end, while the MCU and radio boards are stacked proximally,
connected through compact board-to-board connectors.
This vertical integration approach minimizes lateral footprint, preserves mechanical robustness,
and maintains signal integrity under laboratory testing conditions.
The resulting WIT prototype is integrated into the dedicated compartment shown in Fig.~\ref{fig:compartment}(a). This compartment is subsequently incorporated into the complete oloid-shaped capsule used in the experimental campaign, shown in Fig.~\ref{fig:compartment}(b).

%========================
% IV. EXPERIMENTAL SETUP
%========================
\section{Experimental Validation of the Proposed WIT Prototype}
\label{sec:experimental_setup}

The performance of the proposed WIT module was evaluated
through a series of controlled benchtop experiments.
The objective of the experimental campaign was threefold:
(i) to assess wireless throughput,
(ii) to verify image transmission and reconstruction quality,
and (iii) to quantify subsystem-level power consumption.

All measurements were performed under benchtop conditions
using a colon phantom and an oloid-based magnetic actuation platform
to emulate realistic capsule positioning scenarios.

%------------------------
\subsection{Experimental Setup}

%------------------------
\subsubsection{Colon Phantom and Robotic Manipulation}
\label{subsec:phantom_setup}

The proposed system was integrated into the complete oloid-shaped magnetic capsule described in \cite{Greenidge2025}, as depicted in Fig.~\ref{fig:compartment}(b). In addition to the compartment housing the WIT electronics, the capsule incorporates an internal permanent magnet and was positioned inside the colon phantom. Magnetic actuation was provided by the complete platform, in which an external permanent magnet is mounted at the end effector of a medical-grade, seven-degree-of-freedom KUKA LBR Med R820 robotic manipulator. By varying the pose of the external magnet, the platform applies controlled magnetic forces and torques to translate, orient, and roll the capsule; thus, the KUKA arm is one component of the overall magnetic actuation system rather than the manipulation system itself. An Olympus PCF-160AL EVIS EXERA colonovideoscope (Olympus Corporation, Tokyo, Japan) was also integrated in the capsule and used as the wired high-resolution reference camera for qualitative comparison of the reconstructed images.

During these experiments the capsule-side electronics were tethered to an external regulated power supply, in order to isolate and measure power
consumption of individual subsystems.

%------------------------
\subsubsection{Throughput Measurement Methodology}
\label{subsec:throughput_setup}

Wireless link performance was characterized through \textit{iperf} measurements conducted in UDP mode between the WIT system
configured as station (STA) and an identical SX-SDMAH module connected to a Raspberry~Pi operating as access point (AP).
The Raspberry~Pi acted as the external gateway, performing link establishment, packet reception, data reconstruction, real-time throughput monitoring,
and transmission of bidirectional control commands for exposure time and LED intensity.
The reported metric corresponds to application-layer throughput, thus including protocol overhead, providing a realistic estimate of the effective data rate.

Throughput measurements were performed under both steady and dynamic conditions, with no appreciable differences in the results, confirming the robustness of the wireless link under the tested scenarios.
%------------------------
\subsubsection{Power Measurement Methodology}
\label{subsec:power_setup}

Power consumption was characterized by directly measuring the current drawn by individual subsystems of the WIT module.

Current waveforms were acquired using a digital oscilloscope Tektronix TBS2000 in combination with the high-bandwidth current probe Tektronix TCP0030A. Separate and independent measurements enabled the identification of:
\begin{itemize}
    \item baseline digital consumption,
    \item illumination-dependent power variation,
    \item high-current bursts associated with the RF analog front-end (AFE).
\end{itemize}

Average power was calculated by averaging the measured current waveform over the selected observation interval and multiplying the resulting average current by the supply voltage.
These measurements form the basis for the energy-per-frame and energy-per-bit estimates reported in Section~\ref{sec:results}.
%========================
% V. EXPERIMENTAL RESULTS
%========================
\subsection{Experimental Results}
\label{sec:results}

This section reports the quantitative results obtained from the benchtop validation of the proposed WIT module.
The analysis is structured as follows: (i) wireless link performance, (ii) image transmission assessment, and (iii) power and energy characterization.

%------------------------
\subsubsection{Wireless Link Performance}
\label{subsec:results_link}

Application-layer throughput was measured using \textit{iperf} between the capsule-side IEEE 802.11ah transceiver as STA,
and the Raspberry~Pi-based transceiver as AP.

\begin{figure}[!t]
    \centering
    \includegraphics[width=0.9\columnwidth]{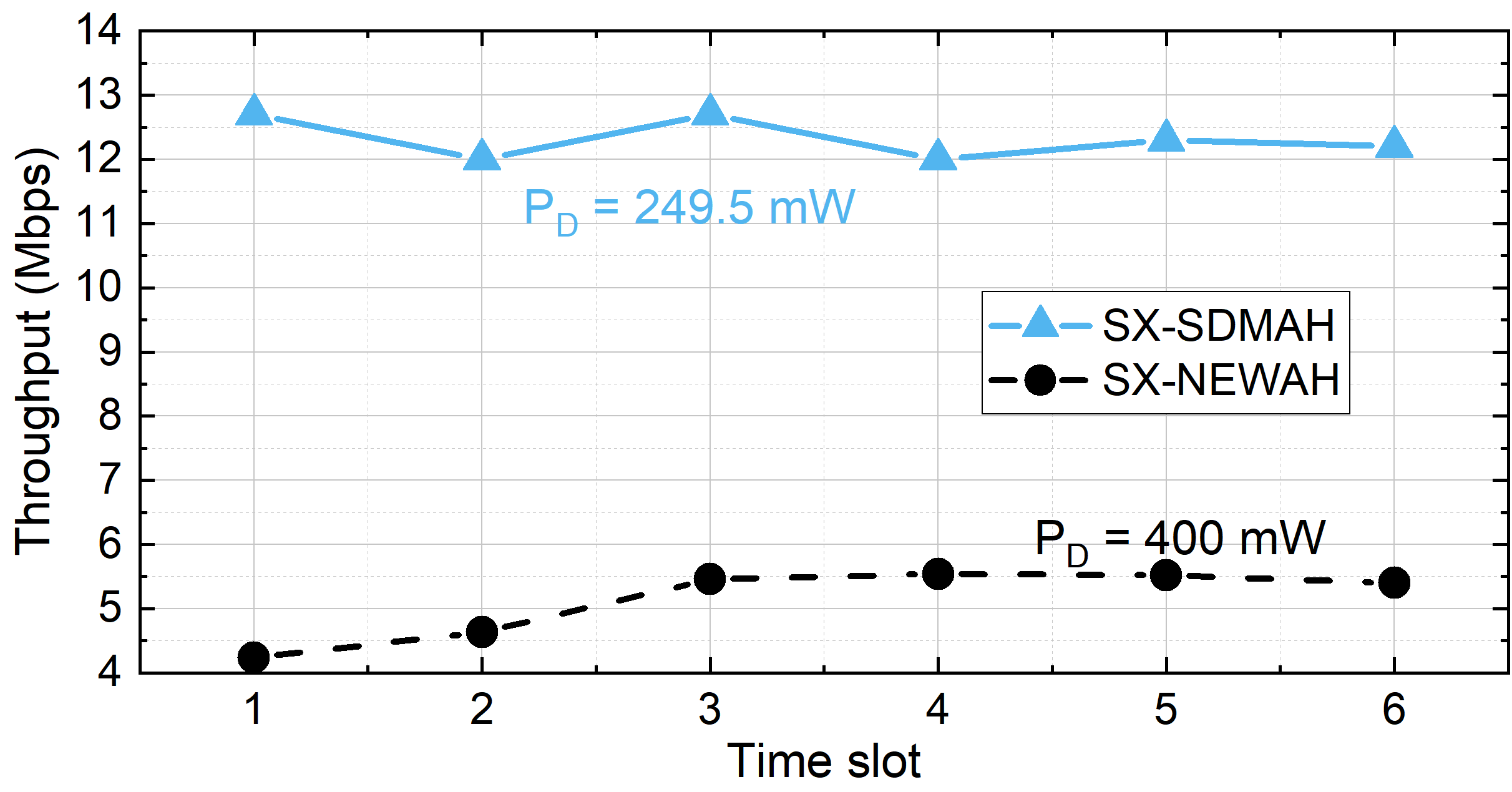}
    \caption{Comparison of the application-layer throughput achieved by the proposed SX-SDMAH-based WIT platform and by the previous SX-NEWAH Wi-Fi HaLow platform. The average dissipated power associated with each hardware is also indicated.}
    \label{fig:comp_wifi_module}
\end{figure}

As shown in Fig.~\ref{fig:comp_wifi_module}, the proposed SX-SDMAH-based WIT platform sustains an application-layer throughput in the 12-13 Mbps range with limited short-term fluctuations, confirming the stability of the IEEE~802.11ah link under the considered laboratory conditions. When compared with the previous SX-NEWAH platform, already employed for the preliminary in-vivo validation discussed in Section~III, the newer module provides a markedly higher throughput, approximately three times larger, while also reducing the dissipated power from about 400 mW to about 250 mW. This comparison highlights the technological improvement enabled by the transition from the earlier NRC7292-based solution to the more recent MM6108-based implementation.

Considering a frame size of $320 \times 320$ pixels with 10-bit effective depth, the number of transmitted bits per frame is:
\begin{equation}
N_{\text{bits}} = 320 \times 320 \times 10 \approx 1 \text{ Mbit}
\label{eq:frame_bits}
\end{equation}
Considering the minimum measured throughput of $DR \approx 12$ Mbps, the theoretical transmission time per frame is:
\begin{equation}
T_{\text{frame}} =
\frac{N_{\text{bits}}}{DR}
\approx 0.083 \text{ s}
\label{eq:frame_time}
\end{equation}
In principle, this bandwidth supports frame rates above 10 fps. In practice, commercial capsule systems typically operate in the range of 1--5 fps,
enabling additional reduction of average power consumption through duty cycling. 

Nevertheless, the availability of higher instantaneous frame rates enables adaptive frame-rate strategies, where the acquisition speed increases during rapid capsule motion.
This improves the sampling density of the gastrointestinal mucosa and reduces the risk of missed detections,
benefiting both physician-based inspection and AI-assisted detection algorithms.
%------------------------
\subsubsection{Image Transmission Performance}
\label{subsec:results_images}
The ability of the WIT module to transmit and reconstruct color frames was evaluated in realistic benchtop scenarios.
\begin{figure}[!t]
    \centering
    \includegraphics[width=\columnwidth]{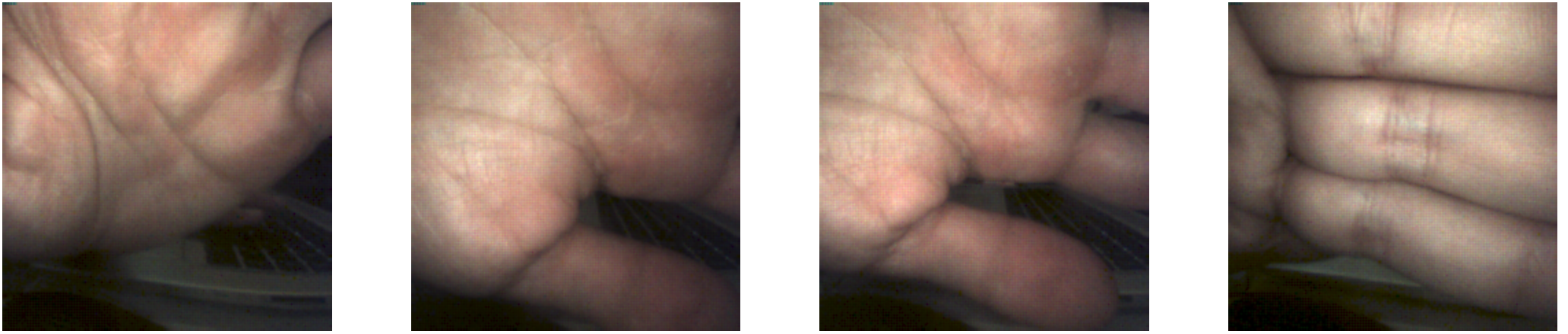}
    \caption{Sequence of $320 \times 320$ color frames wirelessly
    transmitted through the IEEE 802.11ah link.}
    \label{fig:sequenceofhand}
\end{figure}
Fig.~\ref{fig:sequenceofhand} shows a representative sequence of transmitted frames.
Color reconstruction is stable, and no evident packet-induced artifacts are observed.

\begin{figure}[!t]
    \centering
    \includegraphics[width=\columnwidth]{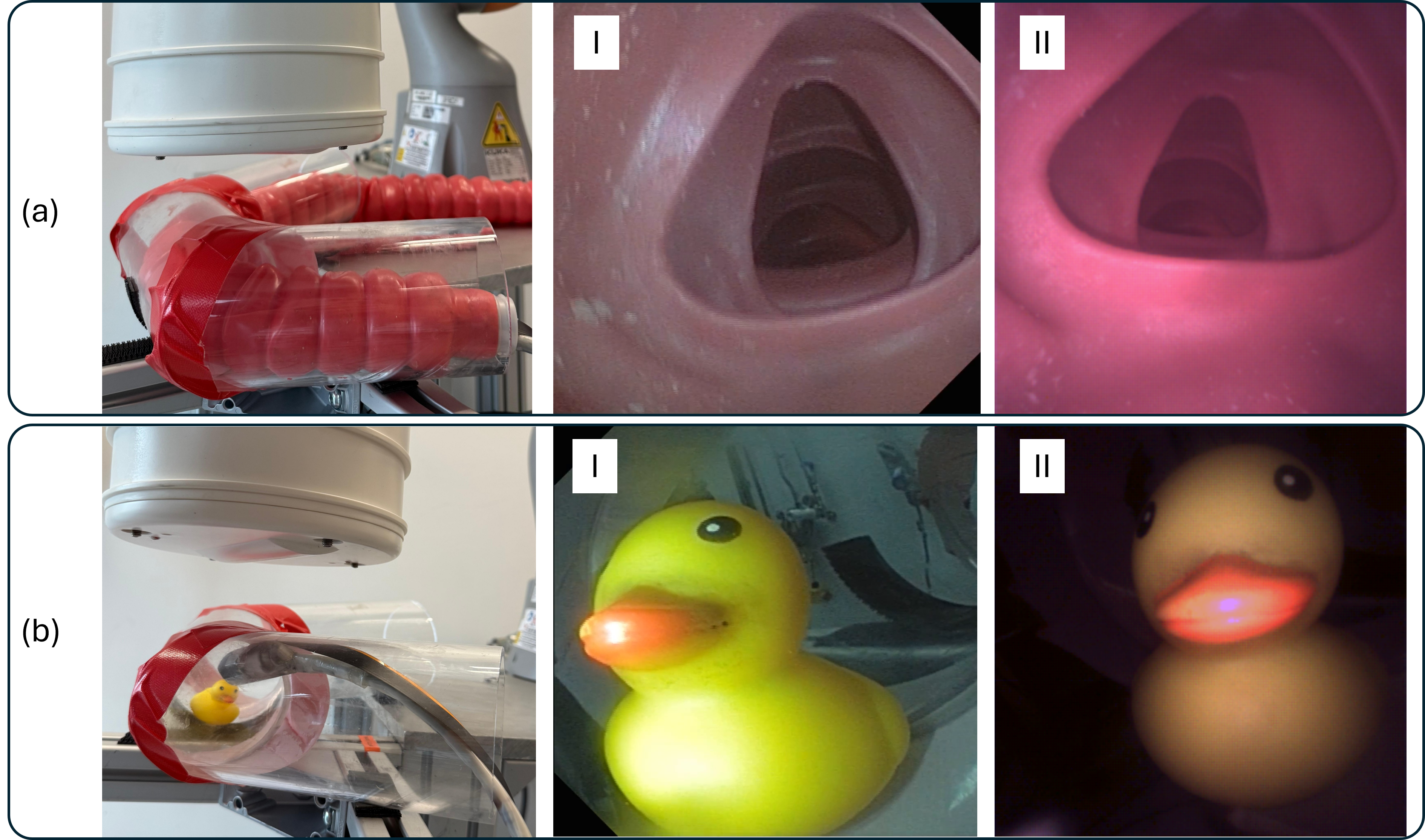}
    \caption{Comparison between the images acquired with the wired Olympus PCF-160AL high-resolution reference colonovideoscope (I) and with the proposed miniaturized WIT module (II) in two representative scenarios (a, b).}
    \label{fig:WIT_Acquisitions}
\end{figure}

A qualitative comparison with the wired Olympus PCF-160AL high-resolution reference colonovideoscope is shown in Fig.~\ref{fig:WIT_Acquisitions}.
As expected, the reference camera provides higher spatial resolution
and finer texture detail.
However, the WIT module delivers full-color images with clearly recognizable structures and contours,
which are adequate for navigation and visual feedback in WCE applications. In all imaging acquisitions, it was not necessary to exceed 60\% LED brightness.
%------------------------
\subsubsection{Power and Energy Characterization}
\label{subsec:results_power}
The power consumption of the WIT subsystem was quantified by directly measuring
the current drawn by each functional block using a current probe and a high-bandwidth oscilloscope.

\begin{figure}[!t]
    \centering
    \includegraphics[width=\columnwidth]{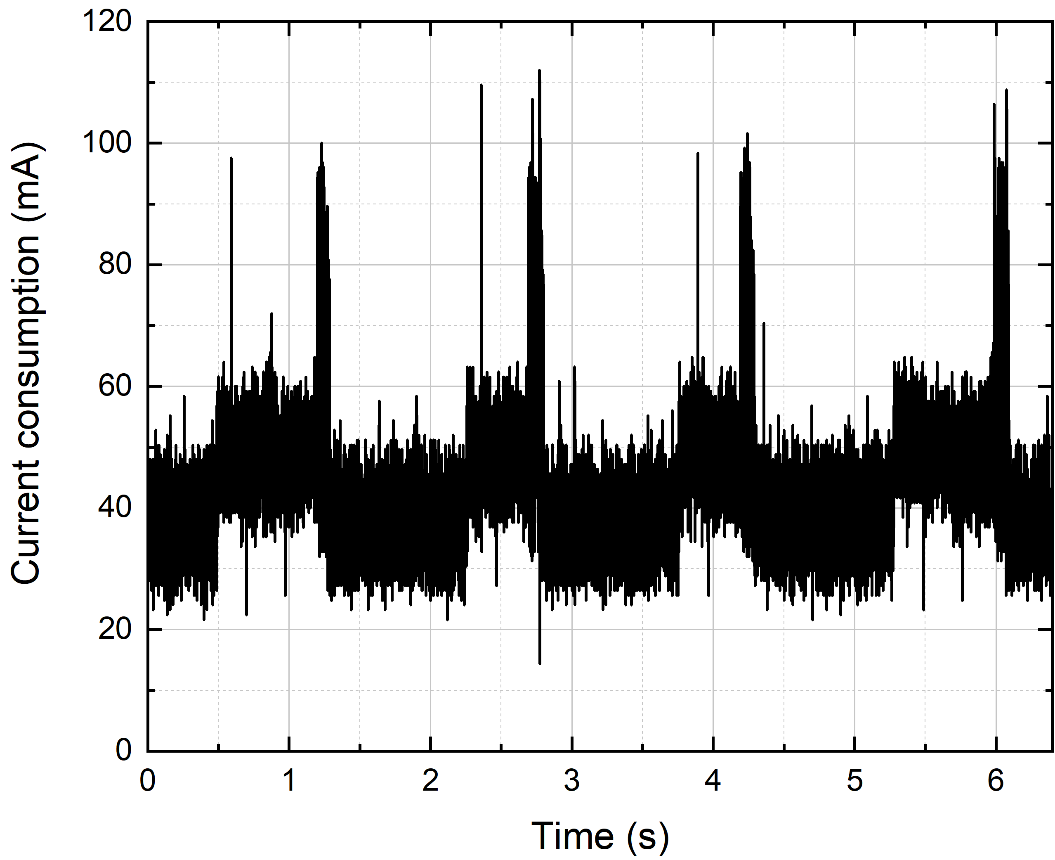}
    \caption{Measured current consumption of STM32U575CIU6Q and SX-SDMAH digital core.}
    \label{fig:plot_MCU_WDT}
\end{figure}
The first set of measurements focused on the baseline consumption of the digital processing chain (microcontroller and the core section of the transceiver).
Both devices are supplied at 3.3 V. As shown in Fig.~\ref{fig:plot_MCU_WDT}, the combined current drawn by the MCU and the HaLow core
is approximately 43 mA:
\begin{equation}
P_{\text{digital}} = V \cdot I
\approx 3.3 \text{ V} \times 43 \text{ mA}
= 142 \text{ mW}
\label{eq:digital_power}
\end{equation}
This value is consistent with those reported in the datasheets, considering that the contributions of the two devices are comparable.
It therefore serves as a reference for extracting the incremental contribution of the remaining subsystems.

The second set of measurements was dedicated to evaluating the power required by the LEDs.
Two operating points were considered: 30\% and 60\% brightness. Since the LEDs are supplied by the same channel that
powers the digital pipeline, the average current drawn by the LEDs alone can be
extracted by subtracting the previously measured baseline. At 60\% brightness, the LEDs consume approximately 23.8 mA at 3.3 V,
corresponding to about 78.9 mW. At 30\% brightness, the current decreases to around 14.7 mA ($\approx$ 48.5 mW).
It is worth noting that exceeding 60\% brightness is not necessary since this level already provided sufficient illumination
for both the colon phantom and the rubber-duck scenes.

The NanEyeC camera itself contributes only marginally to the overall power budget. Direct measurements indicate a current consumption
of approximately 1.5 mA at a supply voltage of 3.3 V, equivalent to almost 5 mW.
\begin{figure}[!t]
    \centering
    \includegraphics[width=\columnwidth]{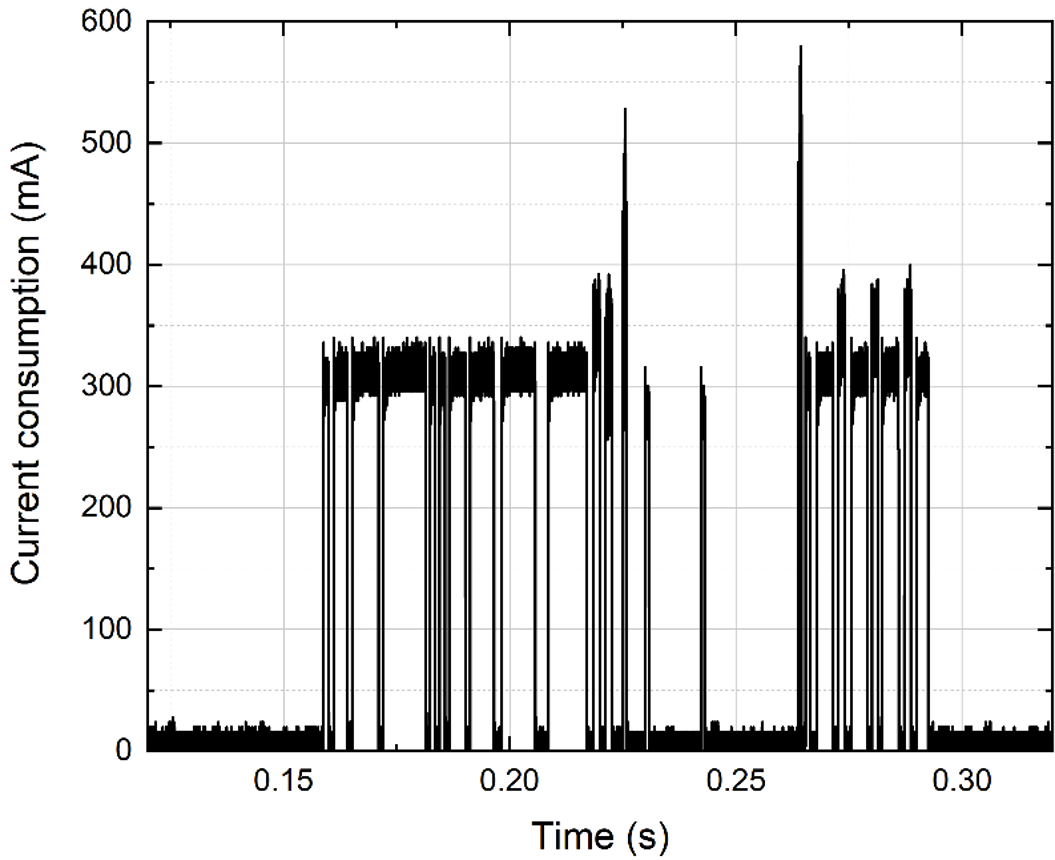}
    \caption{Overall current-consumption profile during image acquisition and RF transmission.}
    \label{fig:plot_consumption_RF}
\end{figure}
The last set of measurements focused on the RF AFE of the transceiver, which is active only during frame transmission.
The current waveform captured during the transmission of a single frame is reported in Fig.~\ref{fig:plot_consumption_RF} and
shows distinct current pulses, with peaks reaching approximately 350 mA and occasional spikes above 500 mA.
By averaging the current profile over a single burst transmission, the effective contribution of the RF AFE corresponds
to an average current of approximately 75.6 mA at 3.3 V.
\begin{equation}
P_{\text{AFE}} = V \cdot I
\approx 3.3 \text{ V} \times 75.6 \text{ mA}
= 249.48 \text{ mW}
\label{eq:afe_power}
\end{equation}
Although the instantaneous peaks are relatively high, their duration is very brief, and the resulting time-averaged power remains
compatible with the constrained power budget.
With a transmission time per frame of $T_{\text{frame}}$, the energy required to transmit a single frame can be estimated as:
\begin{equation}
E_{\text{frame}} =
P_{\text{AFE}} \cdot T_{\text{frame}}
= 249.48 \text{ mW} \times 83 \text{ ms}
= 20.7 \text{ mJ}
\label{eq:energy_frame}
\end{equation}
The corresponding energy per bit is:
\begin{equation}
E_{\text{bit}} =
\frac{E_{\text{frame}}}{N_{\text{bits}}}
= \frac{20.7 \text{ mJ}}{1 \text{ Mbit}}
= 20.7 \text{ nJ/bit}
\label{eq:energy_bit}
\end{equation}

To evaluate the energy consumption of the complete imaging and transmission chain, all subsystem contributions were considered simultaneously. At 60\% LED brightness, the total active power is around 477 mW. Thus, while the RF AFE requires approximately 20.7~nJ per transmitted bit, the complete imaging and wireless transmission chain requires approximately 39.6~nJ/bit.

\begin{comment}

COMPARATIVE ANALYSIS AND DISCUSSION
\label{sec:discussion}

\begin{table}[!t]
\centering
\caption{Comparison with Representative Wireless Capsule Imaging Approaches}
\label{tab:soa}
\begin{tabular}{lccc}
\hline
Approach & Frequency & Throughput & Validation \\
\hline
BLE telemetry \cite{??}        & 2.4 GHz  & $<$1 Mbps   & Bench \\
UWB telemetry \cite{??}         & 3--5 GHz & Multi-Mbps  & Lab \\
This work                             & 900 MHz  & 12--13 Mbps & Phantom \\
\hline
\end{tabular}
\end{table}

\begin{table}[!t]
\centering
\caption{Measured Power Consumption of WIT Subsystems}
\label{tab:power_breakdown}
\begin{tabular}{lccc}
\hline
Component & Current & Supply Voltage & Power \\
\hline
NanEyeC Camera & 1-1.5 mA & 5 V & 5--7.5 mW \\
LEDs (30\%) & 14.7 mA & 3.3 V & 48.5 mW \\
LEDs (60\%) & 23.9 mA & 3.3 V & 78.9 mW \\
MCU + Radio Core & 43 mA & 3.3 V & 142 mW \\
Radio RF AFE & 75.6 mA & 3.3 V & 250 mW \\
\hline
\end{tabular}
\end{table}

\end{comment}
%========================
% VII. CONCLUSIONS
%========================
\section{Conclusions}
\label{sec:conclusions}

This paper presented the design and experimental validation of a miniaturized WIT module for WCE based on a sub-GHz IEEE 802.11ah Wi-Fi HaLow link. Before developing the integrated WIT prototype, the feasibility of the selected wireless technology was assessed in a clinically relevant porcine model. To the best of our knowledge, these experiments represent the first in-vivo demonstration of Wi-Fi HaLow communication through biological tissues for a WCE application.

The proposed architecture integrates a NanEyeC-based WLI front-end, a low-power STM32U575 microcontroller for image acquisition and packetization, and a compact IEEE 802.11ah transceiver within a capsule-compatible stacked PCB implementation. The complete imaging and transmission chain was subsequently validated inside a colon phantom using the oloid-based magnetic actuation platform described in \cite{Greenidge2025}.

The measured application-layer throughput exceeded 12 Mbps, enabling the uncompressed transmission of $320 \times 320$ color frames at frame rates above 10 fps. Subsystem-level power profiling showed an RF transmission energy efficiency of approximately 20.7 nJ/bit. When the camera, digital processing, RF transmission, and illumination at 60\% LED brightness are considered together, the complete WIT chain exhibits an active power consumption of approximately 477 mW, corresponding to approximately 39.6 nJ/bit.

The electromagnetic analysis confirmed that operation around 900 MHz provides greater penetration depth than 2.4 GHz alternatives while retaining multi-megabit communication capability. Together with the in-vivo results, these findings support the suitability of IEEE 802.11ah as a favorable compromise among propagation robustness, throughput, and energy efficiency for in-body wireless imaging.

The footprint and power consumption of the present prototype are partly determined by the level of integration of the adopted commercial transceiver module. The continued introduction of more highly integrated Wi-Fi HaLow SoCs and modules is expected to reduce both board area and power overhead, allowing the proposed architecture to scale toward smaller and more energy-efficient capsule implementations without changing its underlying communication approach.

Although further integration with wireless power transfer and fully autonomous capsule operation remains to be completed, the combined in-vivo communication assessment and laboratory validation of the miniaturized imaging chain demonstrate that IEEE 802.11ah constitutes a technically viable and scalable wireless backbone for next-generation color WCE systems.

\bibliographystyle{IEEEtran}
\bibliography{references}
\end{document}